\documentclass[fleqn,usenatbib]{mnras}

\usepackage{newtxtext,newtxmath}
\usepackage[T1]{fontenc}

\usepackage{graphicx}	
\usepackage{amsmath}	
\usepackage{xcolor}
\usepackage{subfig}

\NewDocumentCommand{\Msun}{o}{
  \IfNoValueTF{#1}
    {\,\mathrm{M}_{\odot}}
    {10^{#1}\,\mathrm{M}_{\odot}}
}

\newcommand{\Lsun}{\,\mathrm{L}_{\odot}}

\newcommand{\rc}{r_\mathrm{c}}
\newcommand{\rt}{r_\mathrm{t}}
\newcommand{\RG}{R_\mathrm{G}}
\newcommand{\rh}{r_\mathrm{h}}
\newcommand{\rhmax}{\rh(\mathrm{max})}
\newcommand{\drhmax}{\delta\rhmax}
\newcommand{\Qstar}{Q_*}
\newcommand{\tDSC}{\tau_\mathrm{DSC}}
\newcommand{\ttDSC}{\widetilde{\tau}_\mathrm{DSC}}
\newcommand{\tC}{\tau_\mathrm{cluster}}
\newcommand{\RBBH}{\overline{R}_\mathrm{BBH}}

\newcommand{\pc}{\,\mathrm{pc}}
\newcommand{\kpc}{\,\mathrm{kpc}}
\newcommand{\Myr}{\,\mathrm{Myr}}
\newcommand{\Myrs}{\,\mathrm{Myrs}}

\newcommand{\Gyrs}{\,\mathrm{Gyrs}}

\newcommand{\secref}[1]{Section~\ref{#1}}
\newcommand{\figref}[1]{Figure~\ref{#1}}
\newcommand{\tabref}[1]{Table~\ref{#1}}

\defcitealias{rostami2024}{Paper~I}

\title[The DSC formation and evolution]{The formation and evolution of dark star clusters  II: The impact of primordial mass segregation}

\author[S. M. Ghasemi et al.]{
S. Mojtaba Ghasemi$^{1}$,
Ali Rostami-Shirazi$^{1}$,  
Pouria Khalaj$^{1}$,
Akram Hasani Zonoozi$^{1}$, and
\newauthor  Hosein Haghi$^{1,2}$\thanks{E-mail: haghi@iasbs.ac.ir}
\\
$^{1}$Department of Physics, Institute for Advanced Studies in Basic Sciences (IASBS), PO Box 11365-9161, Zanjan, Iran\\
$^{2}$School of Astronomy, Institute for Research in Fundamental Sciences (IPM), PO Box 19395 - 5531, Tehran, Iran\\
}

\date{Accepted XXX. Received YYY; in original form ZZZ}

\pubyear{2023}

\begin{document}
\label{firstpage}
\pagerange{\pageref{firstpage}--\pageref{lastpage}}
\maketitle

\begin{abstract}

We investigate the impact of primordial mass segregation on the formation and evolution of dark star clusters (DSCs). Considering a wide range of initial conditions, we conducted $N$-body simulations of globular clusters (GCs) around the Milky Way. In particular, we assume a canonical IMF for all GCs without natal kicks for supernovae remnants, namely neutron stars or black holes. Our results demonstrate that clusters with larger degrees of primordial mass segregation reach their DSC phase earlier and spend a larger fraction of their dissolution time in such a phase, compared to clusters without mass segregation. In primordially segregated clusters, the maximum Galactocentric distance that the clusters can have to enter the DSC phase is almost twice that of the clusters without primordial mass segregation. Primordially segregated clusters evolve with a higher number of stellar mass black holes, accelerating energy creation in their central regions and consequently increasing evaporation rates and cluster sizes during dark phases. The simulations reveal that aggregating heavy components at the centre doubles the time spent in the dark phase. Additionally, the study identifies potential links between simulated dark clusters and initial conditions of Milky Way globular clusters, suggesting some may transition to dark phases before dissolution. Higher primordial mass segregation coefficients amplify the average binary black hole formation rate by 2.5 times, raising higher expectations for gravitational wave emissions.

\end{abstract}


\section{Introduction}
The dynamical evolution of a star cluster is a complex process influenced by a number of internal and external factors. One of the important internal factors is the evolution of massive stars, which can yield the so-called "stellar remnants" such as neutron stars (NSs) or black holes (BHs) \citep{oppenheimer1939continued, iben1984single, iben1991single, fryer1999, heger2003}. These remnants have extremely high mass-to-light ratios ($M/L\gg\Msun/\Lsun$) and therefore are known as "dark stellar remnants". They are mainly the result of supernovae (SNe). The mass ejected by a SN and the associated momentum is not essentially distributed isotropically. This leads to net non-zero momentum transfer to the stellar remnant, called the "natal kick". It was once believed that massive stellar remnants would be ejected from their progenitor clusters due to their natal kicks being larger than the escape speed of the clusters \citep{lyne1994,repetto2012,janka2013}. However, recent studies have shown that a significant number of BHs and NSs can remain in old clusters \citep{Maccarone2007,moody2008,wong2010, Shih2010,Barnard2011,Maccarone2011,strader2012,Chomiuk2013}. The distribution and magnitude of natal kicks for BHs and NSs are still under debate. The retention of these stellar remnants can have important consequences for the dynamical evolution of clusters. They can act as strong gravitational sources that interact with other cluster members and contribute to the formation of binaries and higher-order systems.

Two-body relaxation and dynamical friction in a star cluster cause massive stars to lose their speed and therefore sink towards the centre of the cluster. Subsequently, as a result of striving for energy-equipartition, the low-mass stars gain more speed enabling them to travel farther from the cluster centre. This is the mechanism through which dynamical mass segregation occurs \citep{spitzer2014dynamical,bonnell1998}. Assuming that the remnants' natal kicks are relatively small compared to the escape speed of clusters, a large number of BHs will remain in the cluster. They cannot reach an equilibrium with the low-mass stars via energy equipartition and undergo runaway segregation toward the centre of the cluster. Such clusters are said to be Spitzer unstable \citep{spitzer2014dynamical} and form a gravitationally bound subsystem at their centre, which mainly consists of BHs. Such a subsystem is referred to as a BH sub-cluster (BHSub). 

A BHSub can pump energy into the outer part of a cluster via three mechanisms, i.e. stellar evolution, mass segregation of BHs, and few-body encounters in BHSub (\citealt{breen2013,rostami2024}, hereafter \citetalias{rostami2024}). During the early phases of a cluster's evolution, the mass segregation of BHs transfers energy to low-mass stars through dynamical friction. The segregated BHSub is dynamically active, and many binary BHs (BBHs) form via the three-body encounters \citep{spitzer2014dynamical,Heggie2003}. Through the subsequent encounters between BBHs and single BHs, the binary system becomes harder, while the single BH gets recoiled to larger orbits and transfers its gained kinetic energy to the low-mass stars via two-body encounters \citep{Banerjee2017}. The single BH sinks back to the centre via dynamical friction. This loop continues, with each encounter hardening the BBH and increasing the recoil speed. This eventually causes stars in the cluster outskirts to gain enough speed to escape the cluster. The evaporation rate of stars continuously increases so that in the last stages of cluster evolution, the cluster includes only a few low-mass stars orbiting the central BHSub. Observationally, the remaining low-mass stars appear to be in a super-virial state with a high overall mass-to-light ratio. This is because the velocity dispersion of the remaining stars is enhanced by the invisible BHs. 

When the evaporation time-scale of bright stars from the outer region of the cluster due to the Galactic tidal field is shorter than the self-depletion time-scale of its BHSub, a new kind of star cluster, known as a dark star cluster (DSC), which is dominated by BHs, can be formed. This highly-evolved phase of star clusters was first predicted to exist by \citet{banerjee2011} via numerical simulations. They defined a cluster to be in a DSC state when the (luminous) stars have a virial coefficient of $\Qstar>1$. This can be compared with the results from other studies such as \citetalias{rostami2024}. In addition, \citet{breen2013} focused on the conditions leading to the formation of DSCs depending on the balance between the tidal disruption rate and ejection rate of BHs.

The degree of mass segregation in a stellar system can be quantified by measuring the correlation between the power-law index ($\alpha$) of the mass function ($\mathrm{d}N \propto m^{-\alpha}\mathrm{d}m$) and the so-called concentration parameter $c$. It is defined as $c = \log_{10}{(\rt/\rc)}$, where $\rt$ the tidal radius and $\rc$ is the core radius of the cluster \citep{de2007,baumgardt2008}. The $c$ parameter measures the spatial distribution of stars of different masses. In a strongly mass-segregated system, the most massive stars are concentrated in the central region of the system, while low-mass stars are found predominantly in the outer regions. In contrast, a weakly mass-segregated system exhibits less difference in the distribution of stars of different masses. 

As stated earlier, mass segregation could be dynamical as a result of two-body relaxation and energy equipartition. It can also be the result of the initial state of cluster formation. The latter form is referred to as primordial mass segregation (PMS). PMS implies that in the early stages of a cluster's formation, massive stars tend to be more centrally concentrated than low-mass stars. Although the exact underlying mechanisms that give rise to PMS are not yet fully understood, it is believed that they are closely linked to star formation processes. One such mechanism is turbulence in the star-forming cloud. It can lead to the concentration of gas and dust in the central regions, where the pressure and density are high. This concentration of material can then yield more massive stars, which can grow even further by the accretion of additional material from their surroundings \citep{murray1996,bonnell2002,mcmillan2007}. The degree of the PMS is determined by a coefficient, $S$, which varies from 0.0 to 1.0, representing entirely unsegregated to fully segregated clusters. Lately, there have been observations of several young clusters with large values of $S$ (e.g. \citealt{hillenbrand1997,bonnell1998,fischer1998,de2002,sirianni2002,gouliermis2004,stolte2006,sabbi2008,allison2009,gouliermis2009,de2010, Pavlik2019}).  The photometric and spectroscopic investigations on outer-halo GCs, such as Pal 4 and Pal 14 indicate that they have large half-light radii of $20-30$ pc. Additionally, they show a flat stellar mass function that is depleted in low-mass stars. The centres of these GCs also exhibit clear signs of mass segregation, which is dynamically unexpected since their present-day two-body relaxation times are larger than a Hubble time \citep{Jordi2009, Frank2012, Frank2014}. One way to explain this apparent discrepancy is that these GCs were initially born compact, and the observed mass segregation happened during the early evolution of the clusters \citep{Zonoozi2011, Zonoozi2014, Zonoozi2017, Haghi2015}. 

It is noteworthy to point out that as $S$ increases, the initial expansion of the cluster becomes stronger \citep{vesperini2009PMS}. In particular, \citet{haghi2014} showed that primordially mass-segregated clusters can expand more than their unsegregated identical twin clusters (with the same initial conditions except for $S$) by a factor of about $\approx2$. Moreover, the accumulation of high-mass stars and the resultant stellar remnants at the centre increases the probability of forming a BHSub and consequently a DSC.

In the present paper, we study the effect of PMS on the emergence of the DSC phase and the evolution of its structure for clusters in a MW-like potential. This is achieved by conducting 12 simulations of star clusters with a large PMS coefficient and comparing the simulation outcome with 10 initially non-segregated clusters from \citetalias{rostami2024}. In \secref{method} we elaborate upon our simulation methodology and the adopted initial conditions. \secref{result} embodies the results regarding the evolution of different cluster parameters such as $\Qstar$, half-mass radius ($\rh$), and the duration of the DSC phase ($\tDSC$). In particular, the last two subsections of \secref{result} address the retention of BHs in clusters and the formation rate of BBHs ($\RBBH$). Finally, we conclude the present study in \secref{conclusion}.


\begin{table}
\centering
\caption{Initial conditions of our simulated clusters. The columns from left to right are half-mass radius, the galactic distance (the mean radius of the circular orbit of the cluster), the dissolution time, and the duration of the DSC phase. Clusters without PMS ($S=0$) are from \citetalias{rostami2024}.}
\label{table:all}
\begin{tabular}{lcccr} 
\hline
Model & $\rh[\pc]$ & $\RG[\kpc]$ & $\tC[\Myr]$ & $\tDSC[\Myr]$\\ 
\hline
& & no PMS ($S=0$) & & \\ \\
C1 & 1 & 2 & 1603 & 61\\
C2 & 1 & 3 & 4415 & 1\\
C3 & 3 & 2 & 920 & 130\\
C4 & 3 & 3 & 2376 & 307\\
C5 & 3 & 4 & 4490 & 340\\
C6 & 3 & 6 & 7942 & 554\\
C7 & 3 & 8 & 13093 & 217\\
C8 & 5 & 2 & 506 & 166\\
C9 & 5 & 3 & 1487 & 434\\
C10 & 5 & 8 & 6224 & 828\\
\hline
& & full PMS ($S=1$) & & \\ \\
C11 & 1 & 2 & 1175 & 151\\
C12 & 1 & 3 & 2830 & 232\\
C13 & 1 & 4 & 6225 & 119\\
C14 & 3 & 2 & 674 & 244\\
C15 & 3 & 3 & 1531 & 379\\
C16 & 3 & 4 & 2641 & 587\\
C17 & 3 & 6 & 4872 & 786\\
C18 & 3 & 8 & 6972 & 864\\
C19 & 3 & 12 & 13128 & 763\\
C20 & 5 & 2 & 304 & 170\\
C21 & 5 & 3 & 1075 & 543\\
C22 & 5 & 8 & 4046 & 1307\\
\hline
\end{tabular}
\end{table}

\section{Simulations}\label{method}

We have conducted our simulations using the collisional $N$-body code '\textsc{NBODY7}' \citep{Aarseth2012}, a development of the widely used \textsc{NBODY6} direct $N$-body evolution code \citep{aarseth1999,aarseth2003,nitadori2012}. The state-of-the-art \textsc{NBODY6/7} codes include many physical phenomena, such as the dynamical evolution of the system of stars including collisions and binary formations (via the \citet{kS1965} regularization method) alongside the evolution of single and binary stars \citep{hurley2000}. This family of codes can be run on graphic processing units, which we utilized to significantly reduce the runtime of simulations.

The initial conditions of our simulated clusters have been generated using the \textsc{MCLUSTER} code \citep{kupper2011}. The initial distribution of positions and velocities of stars is given by a \citet{plummer1911} profile. The initial masses of stars are set to follow the \citet{kroupa2001variation} initial mass function (IMF). It is a two-part power-law function as given in Equation~\ref{IMF} with lower and upper stellar mass bounds of $0.07 \Msun$ and $150 \Msun$, respectively.

\begin{equation}
\frac{\mathrm{d}N(m)}{\mathrm{d}m} \propto
\begin{cases}
m^{-1.3} & 0.07 \leq \frac{m}\Msun < 0.5   \\
m^{-2.3} & 0.50 \leq \frac{m}\Msun < 150
\end{cases}
\label{IMF}
\end{equation}

All modelled clusters initially have a total mass of $3\times\Msun[4]$ with a binary fraction of zero. To investigate how different cluster densities in combination with different PMS coefficients affect clusters in reaching the DSC phase, we divide our simulated clusters into three groups with initial half-mass radii of $\rh(\pc)\in\{1, 3, 5\}$. The adopted stellar metallicity is $0.005$ ($\sim$ 0.25 $Z_{\sun}$). The clusters are in a static galactic potential with three components: a central bulge, a disc, and a dark matter halo. We consider a logarithmic dark matter halo to reproduce the gravitational force of the MW galaxy \citep{miholics2014}. The initial orbital velocities of the clusters are set in such a way as to keep them in a circular orbit at their distances of $\RG$, which range from $2$ to $12\kpc$ for different clusters. We found that clusters that reside at larger distances $\RG>12\kpc$ exhibit a very slow rate of evaporation and cannot reach the DSC phase within the Hubble time. The only exception is models with $\rh= 5\pc$ which correspond to the lowest adopted density in our simulations. The clusters dissolve when only a handful of objects are left bound. To maximize the chance of achieving the DSC phase, we ensure that almost all formed NSs and BHs are retained by assuming a natal kick of zero in our models ($\eta = 1$). As for the PMS, we consider $S\in\{0.0, 1.0\}$ (aka, S0 and S1 models, respectively) corresponding to low and high degrees of PMS, respectively. Our primary focus is to see how different PMS coefficients will affect the evolution of clusters. As a result, clusters with the same values of $S$ and $\rh$ are set to have the same initial masses, velocities, and positions for their single stars. The initial parameters of the simulated models are listed in \tabref{table:all}.

\section{Results}\label{result}

\begin{figure}
\centering
\includegraphics[width=\linewidth]{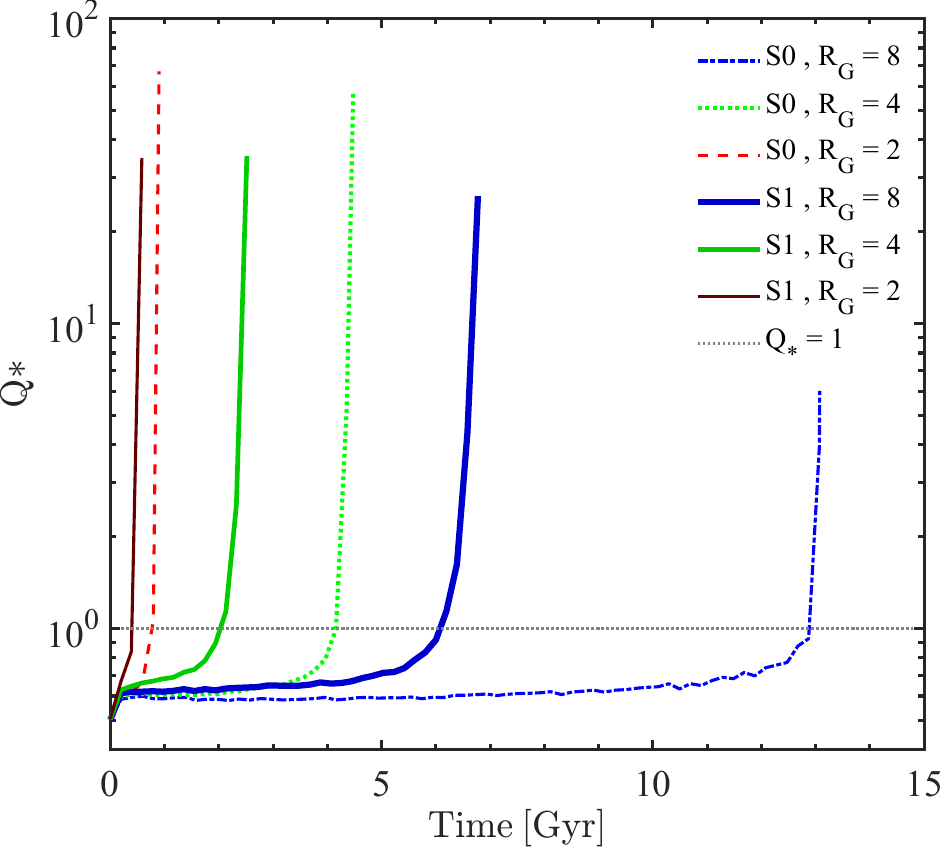}
\caption{The evolution of the virial coefficient for the luminous stars ($\Qstar$) in our simulated clusters from \tabref{table:all}. The vertical axis is logarithmic in scale. Clusters start with $\rh=3\pc$. The solid (dashed) lines correspond to clusters with(out) PMS.}
\label{fig:qs}
\end{figure}

\begin{figure*}
\centering
\includegraphics[width=\textwidth]{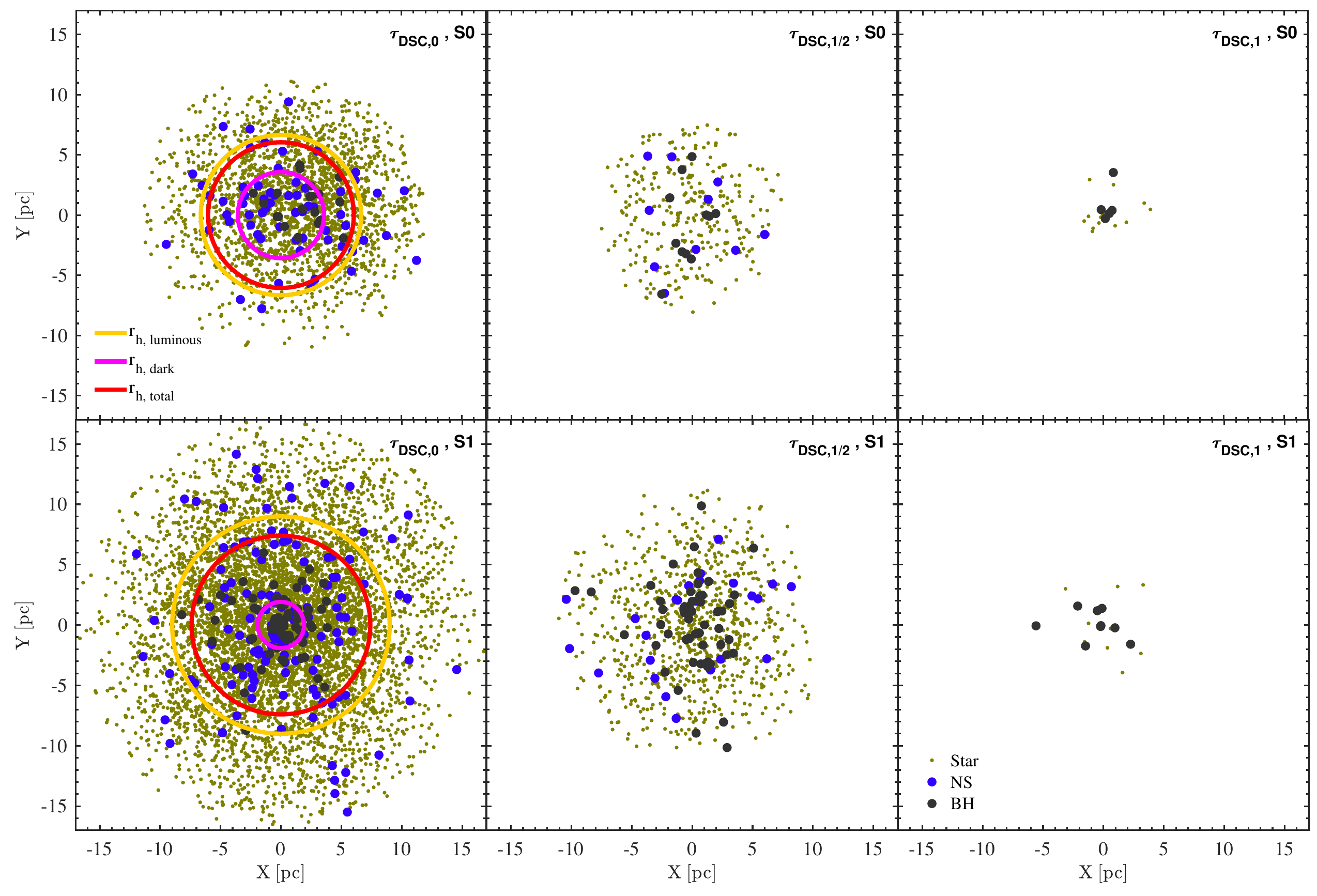}
\caption{The evolution of two (twin) clusters as they reach their DSC phases. Both cluster have $\RG=4\kpc$ and $\rh=3\pc$. The S0 (top panel) and S1 (bottom panel) models are referred to as C5 and C16 in \tabref{table:all}. The panels from left to right show the projected positions of all cluster components for three snapshots. The snapshots correspond to the time that $\Qstar$ exceeds 1.0 ($\tau_\mathrm{DSC, 0}$), when the cluster is halfway through the DSC phase ($\tau_\mathrm{DSC, 1/2}$), and finally when there are still 10 luminous stars inside the cluster ($\tau_\mathrm{DSC, 1}$), right before dissolution. Luminous stars, NSs, and BHs are shown by brown, blue, and black dots, respectively. The red, magenta, and orange circles mark the overall three-dimensional half-mass radii of the clusters, the system of dark (NSs and BHs) components, and luminous stars (including white dwarfs), respectively.}
\label{fig:dscpic}
\end{figure*}     

\subsection{The impact of PMS on DSC structure}

The retention of BHs in a star cluster leads to the transferal of energy to the system of luminous stars via three mechanisms, namely stellar evolution, segregation of BHs, and few-body encounters within the BHSub. These affect non-segregated and primordially segregated clusters differently as elaborated upon below:

\begin{itemize}
\item The early mass loss caused by the stellar evolution of massive stars can make a significant impact on the structure of star clusters \citep{Chernoff1990,Fukushige1995,Takahashi2000,Baumgardt2003,Madrid2012,Kruijssen2014,Krumholz2019}. A newborn star cluster with no PMS has a uniform spatial distribution of massive stars, while its highly segregated twin cluster has a high concentration of massive stars in its central region. Combined with the fact that massive stars ultimately go SNe, this leads to an effective outward flow of energy for clusters with PMS. For the non-segregated case, the energy is uniformly distributed throughout the entire cluster with no effective direction for the flow of energy. 

\item The simulated star clusters are initially in virial equilibrium, i.e. $Q=0.5$. In S0 models, during the first few Myrs of clusters' evolution, mass segregation due to dynamical friction causes the kinetic energy to transfer from massive stars to the system of low-mass stars. This keeps the cluster in virial equilibrium as massive stars and stellar remnants sink toward the centre. In comparison, in S1 models, the initial placement of massive stars in the centre requires low-mass stars to have a higher velocity dispersion initially so that the whole system is in equilibrium. This makes "the evaporation process as a result of energy transfer from the inner to outer layers of the cluster" more efficient and faster. In other words, the low-mass stars are already in a hot state and can leave the cluster with small amounts of energy gained from inner regions. In contrast, S0 models need some time for the segregation of massive stars and transfer of energy to outer regions, making the evaporation process slower than S1 models.  

\item The BHSub is more compact in S1 models, this increases the likelihood of three-body encounters within the core and results in more energy being transferred to the surrounding system of stars.
\end{itemize}

\figref{fig:qs} illustrates the evolution of $\Qstar$ for clusters with and without PMS at different $\RG$ distances of 2, 4, and 8 $\kpc$, with PMS coefficients of $S=0$ and $S=1$. In our simulations, we assume the DSC phase for a cluster begins when the virial coefficient of the system of luminous stars exceeds 1.0 (c.f. \citealt{banerjee2011}). This has been marked by the grey dotted line in the figure ($\Qstar=1.0$), which borders the pre-DSC and the DSC phases. Clusters exhibit a gradual increase in their values of $\Qstar$ in the pre-DSC phase, followed by a rapid change (almost vertical) once they cross the $\Qstar=1.0$ line. In particular, within a period of a few Myrs, the virial coefficient magnifies by a factor of $\ge2$ just before the cluster entirely dissolves. This rapid change primarily occurs after the cluster loses most of its luminous stars while retaining a significant amount of its dark stellar remnants.  

The stronger energy generation of BHSubs in S1 models results in a more pronounced emergence of the DSC phase in these clusters. Unsurprisingly, \figref{fig:qs} also indicates that the orbital radius of a cluster has a significant impact on its mass loss and dissolution rate. Clusters located further away from the centre of the host galaxy experience a weaker potential, resulting in a slower rate of mass loss and a longer dissolution time compared to those located closer to the centre. This can be simply explained as follows. Larger values of $\RG$ lead to an increase in the escape velocity of the cluster, which in turn lowers the evaporation rate of stars and aids in retaining the system of luminous stars for a longer time. As a result, $\Qstar$ exceeds $1.0$ at a later time, and so does the DSC phase.

\figref{fig:dscpic} depicts three snapshots (from left to right in succession) of two clusters with $\RG= 4\kpc$ and an initial $\rh$ of $3\pc$ as they reach the DSC phase. The top and bottom panels correspond to S0 and S1 models, respectively. The S0 model reaches the DSC phase after $4.2\Gyrs$, with a total mass of $M\approx1.8\times\Msun[3]$ and a half-mass radius of $\rh\approx6.1\pc$. In comparison, this occurs after $2.1\Gyrs$ for its twin (S1) at which point its total mass and half-mass radius is $M\approx5.1\times\Msun[3]$ and $\rh\approx7.4\pc$, respectively. Despite their almost equal $Q*$ values, the models differ significantly in the number of their stars, with the S0 and S1 models having approximately 2300 and 6000 luminous objects, and 90 and 200 dark remnants, respectively. The fact that the S1 model has more objects in the DSC phase can be explained as follows. A large number of massive dark stellar remnants in this model reside compactly inside a half-mass radius of $\approx1.9\pc$ which is almost one-third of the cluster's $\rh$. This creates a dense and massive cluster core capable of sustaining a larger number of stars in a super-virial state. In contrast, the dark components of the S0 cluster have $\rh\approx3.6\pc$, which is $\approx0.6$ of the overall $\rh$. Moreover, S1 reaches its DSC phase at an earlier time (relative to its total dissolution time). Therefore, S1 clusters have a higher probability of having more compact sub-systems of massive dark remnants in their core, increasing the likelihood of creating BH-BH or BH-NS binaries or multiples in the cluster.

\begin{figure}
\centering
\centering 
\includegraphics[width=\linewidth]{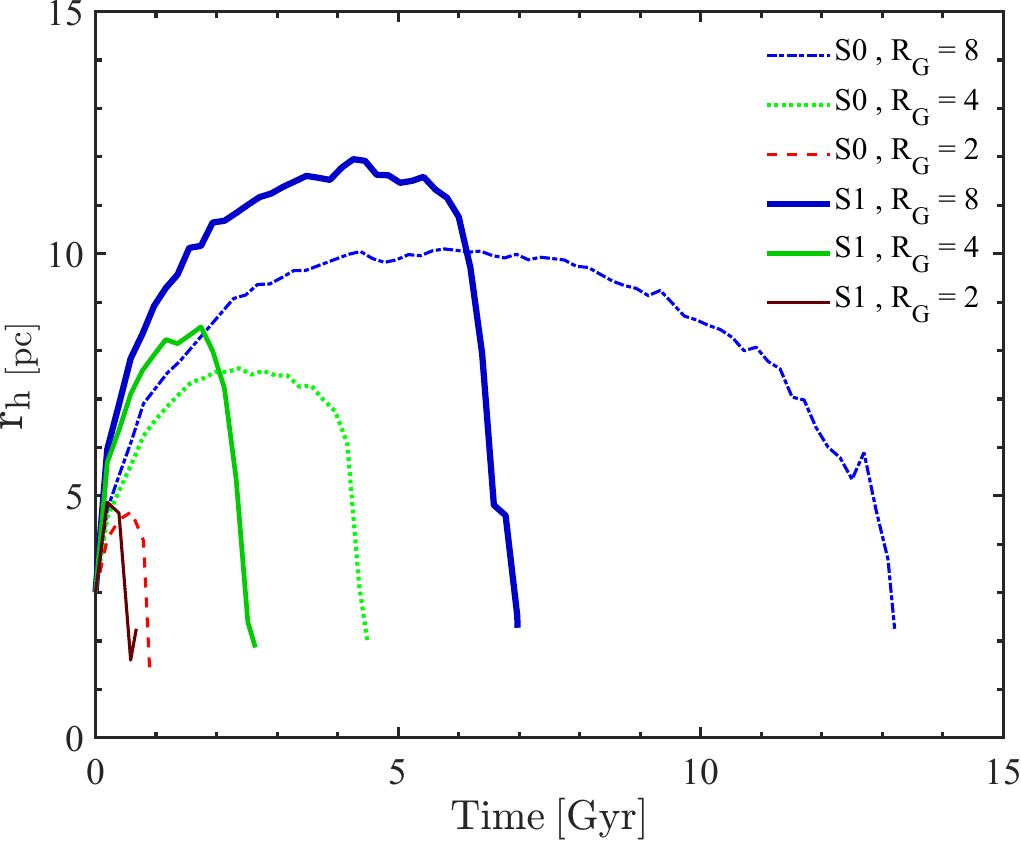}
\caption{The evolution of half-mass radii of six models from \tabref{table:all} all starting with an initial value of $\rh=3\pc$. The line colours are the same as ~\figref{fig:qs}.}
\label{fig:rh}
\end{figure}

\subsubsection{Energy flux and size evolution}\label{size}

Transferal of energy to the system of low-mass and luminous stars during the early stages of a cluster's evolution is mostly dominated by the powerful shocks from SNe and the mass segregation of BHs as a result of low natal kicks of the stellar remnants in our models. At later stages, however, this is driven by the few-body encounters inside the cluster. The transferred energy once reaches the cluster's outskirt, is absorbed by low-mass stars, resulting in speeding them up even further and contributing to the overall expansion of the cluster. 

\figref{fig:rh} demonstrates that S1 models tend to dissolve and expand at a faster rate than their S0 twins. Moreover, the peak of $\rh$ is attained at larger values for S1 models. We define the relative difference in $\rhmax$ as below,
\begin{equation}
\drhmax=\frac{\rh(\mathrm{max, S1})}{\rh(\mathrm{max, S0})}-1
\end{equation}
One observes that $\drhmax$ increases as a function of $\RG$, i.e. $\drhmax\approx0.05$ for $\RG=2\kpc$, $\drhmax\approx0.10$ for $\RG=4\kpc$, and $\drhmax\approx0.18$ for $\RG=8\kpc$. 

One can compare these results with those reported by \citet{haghi2014} regarding the behaviour of $\rh$ as a function of PMS coefficient $S$, for the clusters with no initial BH retention ($\eta = 0$). In particular, when comparing the expansion of the S0 and S1 models, it is evident that models with $\eta = 1$ show less difference in expansion between their S0 and S1 twin clusters than simulated clusters with $\eta = 0$ ($\delta\rh(\mathrm{max;\eta = 1}) < \delta\rh(\mathrm{max;\eta = 0})$). \citet{haghi2014} demonstrated that in models without BH retention evolving at Galactocentric distances of 4 kpc and 8 kpc, the relative difference in $r_h (max)$ for models with and without PMS is about $\drhmax=$0.26 and 0.33, respectively. These values are, on average, two times larger than what is obtained for models with BH retention.

The significant expansion of the mass-segregated clusters compared to the non-mass-segregated clusters is caused by two factors. Firstly, this expansion is due to the high initial speeds of low-mass stars in clusters with large $S$ values (as the cluster starts in a virial equilibrium). Secondly, in models with large S values, the majority of low-mass stars are located in the outer skirt of the cluster. This expansion is more significant in clusters with no initial BH retention, resulting in the loss of BHs, which decreases the gravitational potential of the systems. In the absence of a natal kick, the energy transferred from the BHSub to the low-mass stellar populations in S0 models aids in marginalizing the difference in $\rhmax$ between S0 and S1 models.

\begin{figure}
\centering
\includegraphics[scale=0.575]{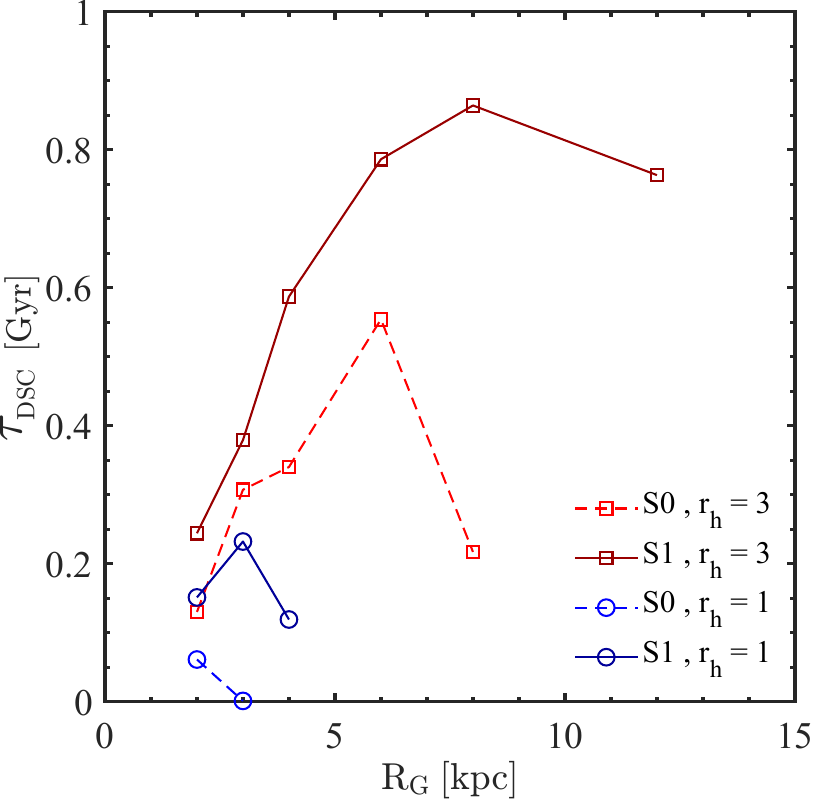}
\includegraphics[scale=0.58]{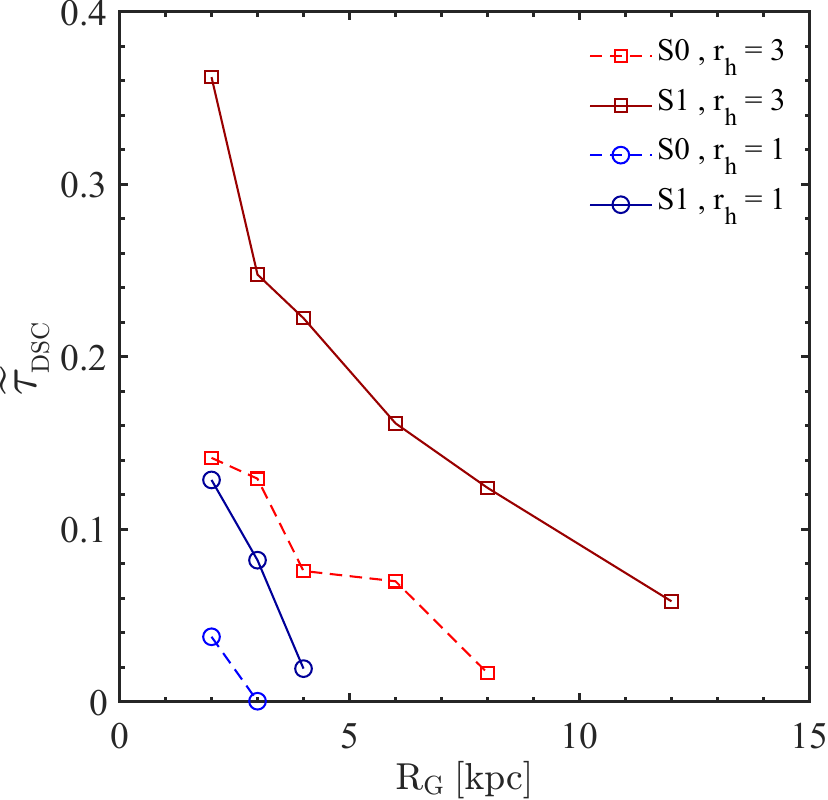}
\caption{Top: the duration of the DSC phase simulated clusters with $\rh(\pc)\in\{1, 3\}$ from \tabref{table:all} as a function of their Galactocentric distances. Bottom: Evolution of the ratio of the DSC phase duration to the dissolution time of clusters utilized in the top panel as a function of $\RG$. Solid and dashed lines correspond to clusters with and without PMS, respectively. Clusters with an initial $\rh$ of 1 and $3\pc$ are depicted by circle and square markers, respectively.}
\label{fig:lifetime}
\end{figure}

\begin{figure*}
\centering
\subfloat \centering {\includegraphics[scale=0.44]{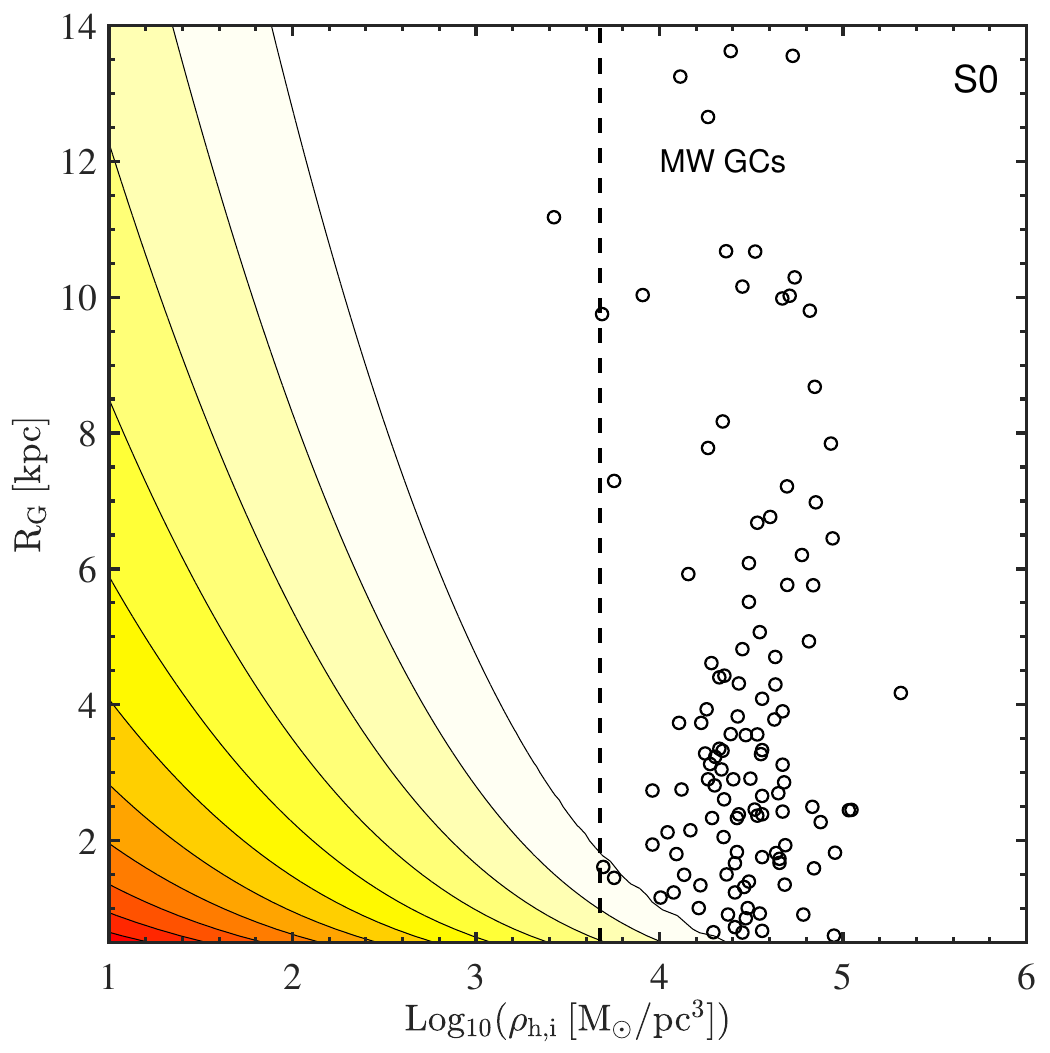}}
\subfloat \centering {\includegraphics[scale=0.44]{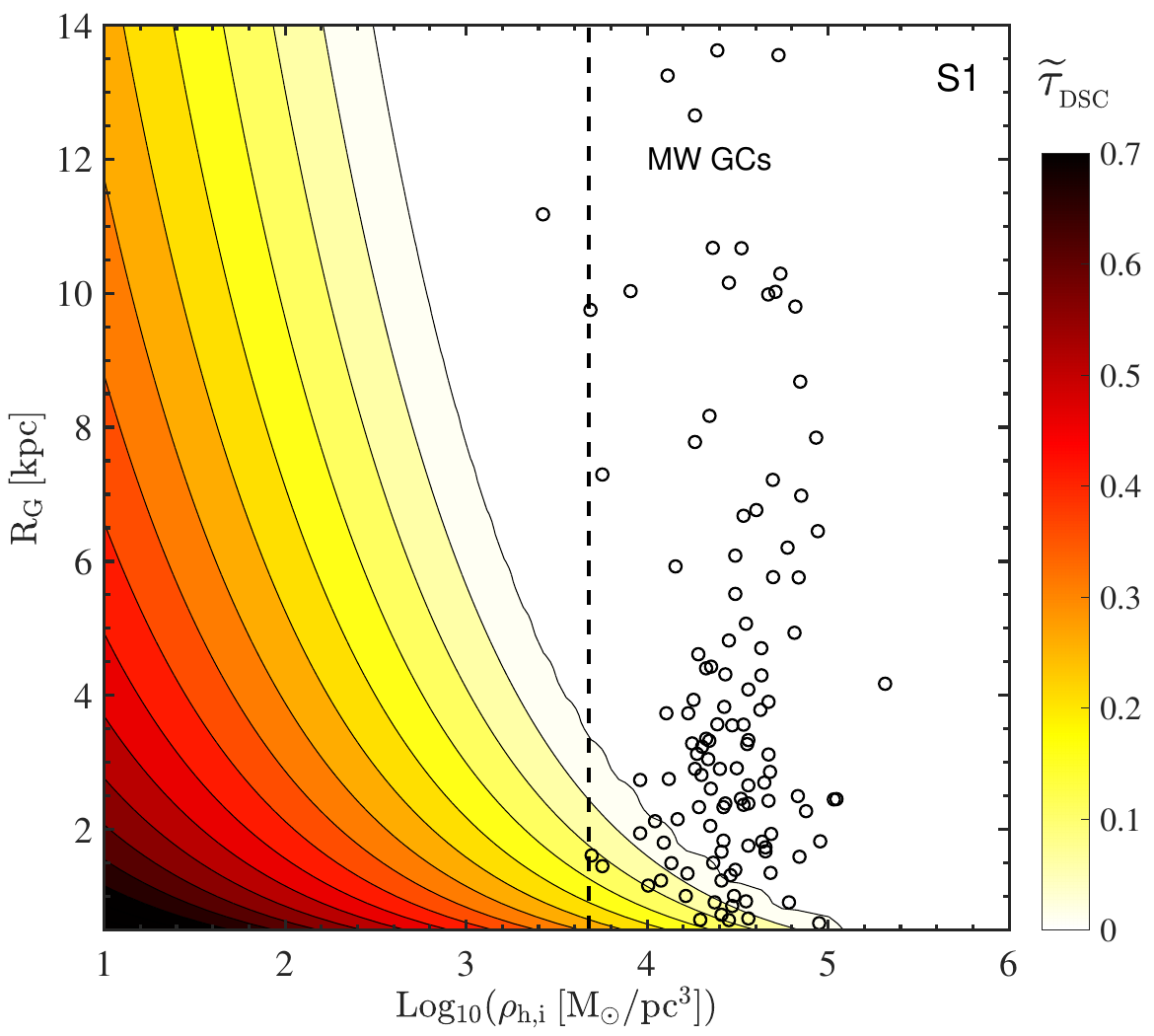}}
\caption{The contours of $\ttDSC$ as a function of $\log_{10}(\rho)$ and $\RG$ for S0 (left) and S1 (right) models. The regions to the right of the black dashed lines enclose the estimated location of MW GCs (black circles) in the given parameter space. The initial densities and half-mass radii of MW GCs are estimated based on \citet{baumgardt2019} and \citet{markskroupa}.}
\label{fig:contour-tdsc-Tdis}
\end{figure*}     

\subsubsection{Duration of the DSC phase} \label{DSCduration}

As mentioned earlier, S1 models experience a higher rate of cluster evaporation and subsequently a shorter dissolution time. \tabref{table:all} lists the dissolution time ($\tC$) of all simulated clusters and the duration of their dark phases ($\tau_\mathrm{DSC}$). It can be seen that S0 models can survive up to 2 times longer than their twin S1 models.

The top panel of \figref{fig:lifetime} provides valuable insight into the impact of the PMS coefficient on the duration of the DSC phase as a function of $\RG$, for clusters with initial $\rh\in\{1,3\}\pc$. It is evident that higher degrees of PMS lead to an earlier emergence of the DSC phase and prolongation of $\tDSC$ for clusters with the same $\RG$. Moreover, we note that the trend of $\tDSC$ versus $\RG$ is not monotonic, but rather, has a turnover point that corresponds to its global maximum. In \citetalias{rostami2024}, it was found that the existence of such a turnover point can be explained by the balance between the evaporation rate of luminous stars and the self-depletion rate of the BHSub. As we move further away from the centre of the Galaxy, the self-depletion time scale of the BHSub becomes shorter than the evaporation time scale of the luminous stars. This results in a decrease in the value of $\tDSC$ as we move towards the outskirt of the Galaxy until it finally reaches zero. \citetalias{rostami2024} also noted that the turnover point is attained at $\RG=2\kpc$ and $\RG=6\kpc$ for clusters with $\rh=1\pc$ and $\rh=3\pc$, respectively. As the PMS coefficient increases, the location of the turnover point moves towards larger orbital radii. This is due to the enhanced energy generation by the clusters' BHSub. In particular, S1 models exhibit this turnover point at $\RG=3\kpc$ ($\rh=1\pc$) and $\RG=8\kpc$ ($\rh=3\pc$).

\citetalias{rostami2024} defined a parameter to quantify the effect of the BHSub on the evolution of star clusters. It is the ratio of the DSC phase duration to the cluster's dissolution time ($\ttDSC=\tDSC/\tC$) and is referred to as the "scaled DSC lifetime". If $\ttDSC=1$, it means that the cluster stays in its DSC phase for its entire lifetime. Conversely, if a cluster never enters the DSC phase, the value of this parameter is zero. As shown in the bottom panel of \figref{fig:lifetime}, $\ttDSC$ decreases significantly as a function of $\RG$. Interestingly, S1 Models on average have $\ttDSC$ values twice as large as those of S0 models. 

\figref{fig:contour-tdsc-Tdis} shows the contour plots of $\ttDSC$ in the parameter space of $\RG$ and the initial density within the half-mass radius ($\rho_\mathrm{h,i}$), for S0 and S1 models. As shown in the figure, the rate of change along each of the axes is negative while holding other variables constant. As expected, $\ttDSC$ reaches larger values in S1 models. This can be more clearly seen when comparing the lower left regions in both panels, i.e. clusters with very low initial densities ($\approx10\Msun\pc^{-3}$) and small values $\RG$. In particular, S1 models in such regions can spend up to 70 per\,cent of their lifetime in the DSC phase, compared to only 35 per\,cent for S0 models. The upper limit on the initial density of clusters to ultimately enter the DSC phase is approximately $1.6\times10^4$ and $10^5$ $\Msun\pc^{-3}$ for S0 and S1 clusters, respectively. This phase does not happen for $\RG$ values larger than $17\kpc$ for S0, and $48\kpc$ for S1 models. 

\figref{fig:contour-tdsc-Tdis} also shows the mean Galactocentric distances and the expected initial densities of 154 MW GCs \citep{baumgardt2019}. On average, the MW GCs have lost about $75$ per\,cent of their initial stellar masses through stellar evolution or long-time dynamical evolution \citep{webb2015back, baumgardt2019}. Our simulations do not include primordial gas expulsion which can affect the densities and radii of clusters (e.g. \citealt{Khalaj2015, Khalaj2016}). We compensate for this effect when inferring the initial characteristics of MW GCs as follows. We assume that the initial stellar masses of clusters were 4 times as large as the present-day dynamical mass of the clusters \citep{Baumgardt2018}. The initial half-mass radii are then estimated using the results from \cite{markskroupa} as further elaborated upon. Since our simulations address the evolution of clusters after the end of gas expulsion, the initial half-mass radii should be multiplied by the factor of $\approx3$. This is essentially the expansion factor for gas clumps assuming the widely adopted star formation efficiency of 0.33 \citep{baumgardt2007}. By applying this expansion factor of 3, we estimate a density of a $\approx3\times\Msun[4]\pc^{-3}$ as the lower limit for the initial density of the MW GCs, marked by the black dashed line in \figref{fig:contour-tdsc-Tdis}. We emphasize that the initial density of MW GCs is on average larger than this value as they are more massive than our modelled clusters.

As pointed out in \citetalias{rostami2024}, MW GCs cannot enter the DSC phase if they had a canonical IMF without high degrees of PMS. The left panel of \figref{fig:contour-tdsc-Tdis} reasserts this claim as almost none of the MW GCs intersect non-zero contour levels of $\ttDSC$. Intriguingly, the right panel of \figref{fig:contour-tdsc-Tdis} shows that the MW GCs placed in the innermost parts of the MW ($\RG\lesssim 2\kpc$) can evolve into and stay in the DSC phase for up to 15 per\,cent of their lifetime if they highly mass segregated at their birth time. As a result, we do not expect to observe any GCs in the DSC phase at large orbital radii unless they had a top-heavy IMF (\citetalias{rostami2024}). Using the right panel of \figref{fig:contour-tdsc-Tdis}, we estimate that up to 14 per\,cent of the MW GCs can evolve into the DSC phase under the assumption of high $S$.

\begin{figure}
\centering 
\includegraphics[width=\linewidth]{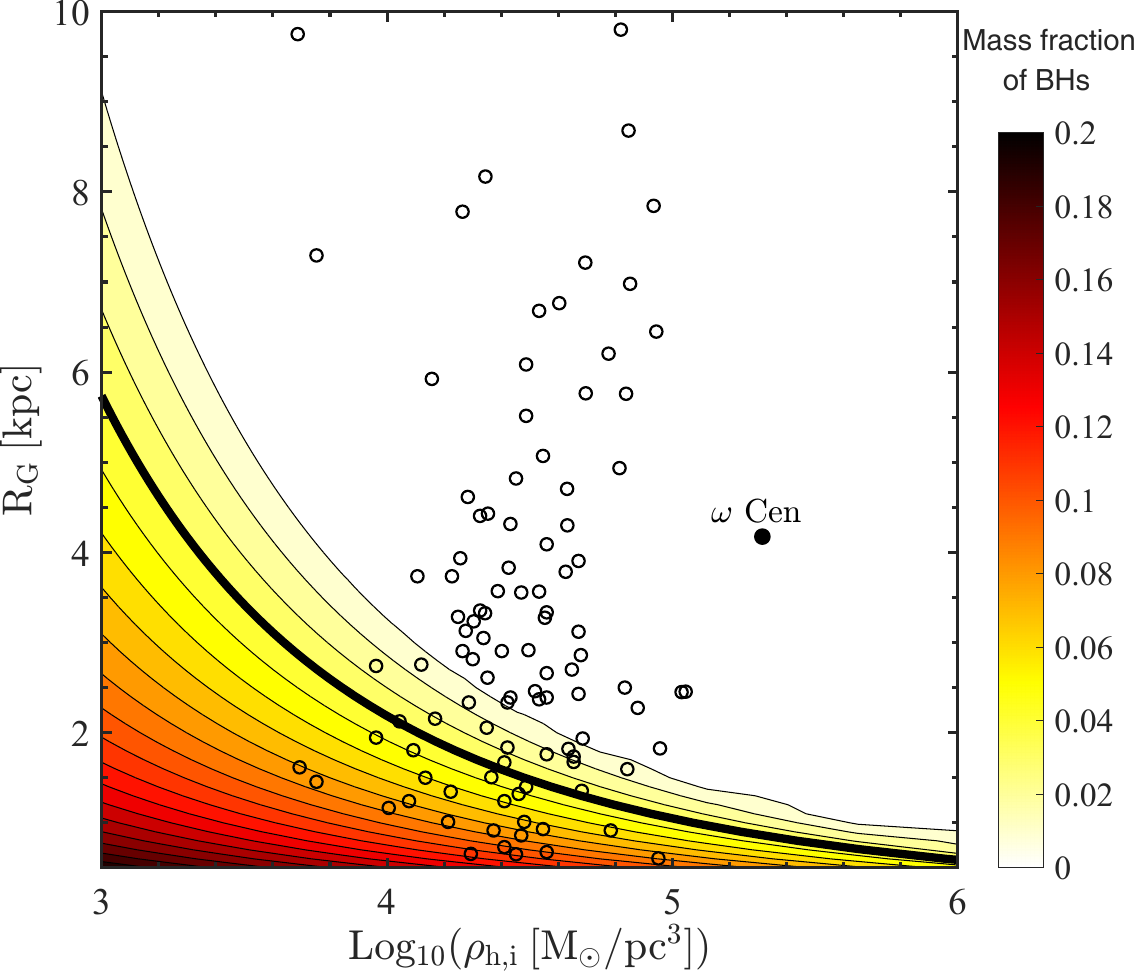}
\caption{Mass fraction of BHs for our S1 clusters after losing 75 per\,cent of their initial mass. The contour plot is a function of the initial orbital radii and initial densities, similar to \figref{fig:contour-tdsc-Tdis}. The thick black curve marks the 4 per\,cent mass fraction. The black circles show the estimated initial parameters of the MW GCs. The filled black circle corresponds to $\omega\,\mathrm{Cen}$.}
\label{fig:BHfraction}
\end{figure}

\subsection{The impact of PMS on the population of retained BHs}

The lack of natal kicks in our models results in a mass fraction of $\approx 4$ per\,cent for BHs in the first few $\Myrs$ of the clusters' evolution. This value can be compared to the initial fraction of BHs in clusters, assuming they had a canonical IMF and a natal kick of zero for remnants. Clusters that reach the DSC phase, experience a sustained growth in their BH mass fraction, so an evolved cluster with a BH mass fraction larger than 4 per\,cent can potentially become a DSC. \figref{fig:BHfraction} shows the mass fraction of BHs for S1 models when they have lost 75 per\,cent of their initial mass (as is the case for the MW GCs). The thick black curve line shows the BH mass fraction of 4 per\,cent. Highly mass-segregated clusters with low initial densities, regardless of their orbital radii, are expected to harbour several stellar-mass BHs until they have lost 75 per\,cent of their initial masses. This is also the case for high-density clusters, albeit only for those at smaller orbital radii. By superimposing the MW GCs (black circles) in the parameter space of \figref{fig:BHfraction}, we find 37 clusters (24 per\,cent) with a population of BHs, and mean Galactic distances of $\leq4\kpc$, of these, 23 have a BH mass fraction of more than 4 per\,cent. These GCs can potentially evolve into a DSC. This is in agreement with our results from \secref{DSCduration} where 14 per\,cent of the MW GCs were found to be able to become DSCs. According to our results, we estimate the present-day population of MW GCs to have BH mass fractions of up to $10$ per\,cent if they have initially been highly mass-segregated. Our results from \citetalias{rostami2024} regarding the full retention fraction of BHs in the MW GCs with a canonical IMF and a negligible $S$ factor, about 13 per\,cent of the MW GCs with low initial densities and small semi-major axes can potentially harbour BHs even after losing three-quarters of their initial masses. This percentage doubles for clusters with exceptionally high $S$ values.

Several recent studies show that GCs such as $\omega\,\mathrm{Cen}$, Pal5, and M22 potentially harbour a considerable number of stellar-mass BHs. Among which, $\omega\,\mathrm{Cen}$ and Pal5 are determined to have BH mass fractions of 4.6 (\citealt{baumgardt2019b}; also see \citealt{zocchi2019}) and 20 \citep{gieles2021} per\,cent, respectively. Moreover, M22 is estimated to have about 5 to 100 BHs \citep{strader2012}. If we assume that GCs form in small half-mass radii, as suggested by \citet{markskroupa}, our results indicate that achieving such BH mass fractions does not suit high-density GCs such as $\omega\,\mathrm{Cen}$ or GCs with mean Galactocentric distances larger than 4 kpc such as M22 or Pal5 if they had a canonical IMF, even with high degrees of PMS. The mass fraction of BHs in these clusters can be justified by assuming a top-heavy IMF, as per our results in \citetalias{rostami2024}, or by adopting a larger initial half-mass radius to enhance the capacity to retain BHs. In conclusion, the present-day retention fraction of BHs for MW GCs with a canonical IMF is limited to 10 per\,cent.

\subsection{The impact of PMS on the dynamical formation of BBHs}

The system of BHs in a cluster with a high PMS coefficient is more centrally concentrated. A denser system of BHs leads to more few-body encounters inside the cluster and will subsequently increase the dynamical formation of BBHs. A higher formation rate for BBHs increases the probability of observing BH X-ray binaries and also increases the possibility of BH-BH mergers, which are potential sources for the emission of strong gravitational waves (GWs) \citep{banerjee2010stellar,morscher2013,abbott2016,Antonini2016,LIGO2017,Askar2017,Rodriguez2018,Samsing2018,Samsing2020,Kremer2019,Hong2020,Weatherford2021,Tiwari2024}. 

\begin{figure}
\centering 
\includegraphics[width=\linewidth]{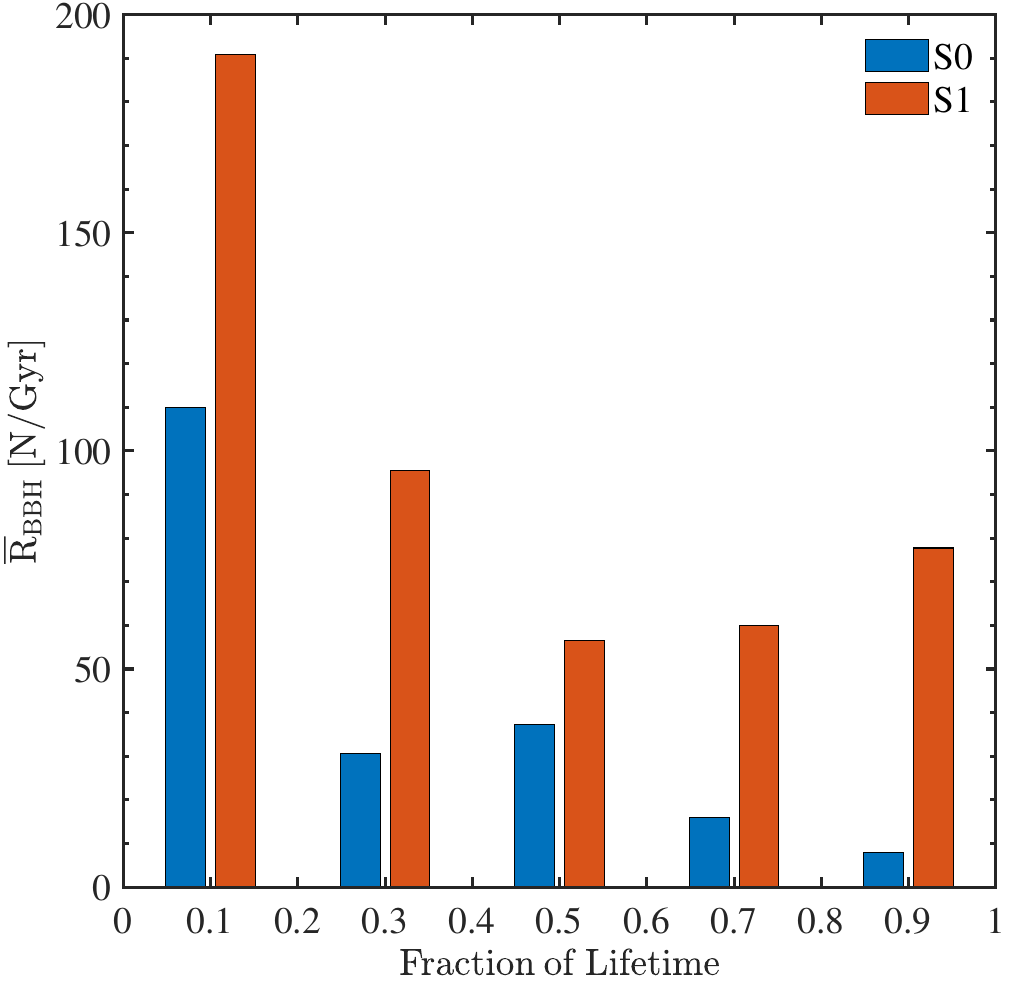}
\caption{Mean binary black hole creation rate at different fractions of clusters' lifetime. The two clusters represent C2 and C12 models that only differ in their PMS coefficients, namely S0 (blue) and S1 (red).}
\label{fig:BBH}
\end{figure}

Our simulations indicate that the dynamical formation of BBHs occurs throughout the lifetime of a star cluster, which is expected to evolve into the DSC phase. We studied the formation rate of BBHs ($\RBBH$) in two simulated models with the same initial conditions but different degrees of PMS: C2 with $S=0$ and C12 with $S=1$. The initial densities of these models are similar to the MW GCs. \figref{fig:BBH} illustrates the evolution of $\RBBH$.
In model C2, $\RBBH$ slows down as the cluster ages and the system of BHs gets depleted. A high degree of initial mass segregation leads to the formation of a larger number of BBHs in the C12 cluster, by a factor of 2 and 10 during the first and last quintiles of the cluster’s lifetime, respectively. On average, the number of dynamically formed BBHs over the lifetime of a star cluster is $2.5$ times larger if it is initially segregated. This is because BHs are more likely to encounter each other in the dense central part of a cluster, which BHs dominate in a segregated cluster. According to the figure, there seems to be an almost monotonic depletion of BBHs for S0 over time, whereas S1 levels off at a constant rate. Based on our assumptions of negligible remnant natal kicks and a canonical IMF for MW GCs placed within mean Galactic orbital radii of $\leq4\kpc$  (so that they can retain some stellar-mass BHs until the present time) we anticipate an overall tenfold boost in BBHs for $S=1$ compared to $S=0$. Similarly, the rate of X-ray binary creation and GW emission can also increase in response to an increase in the PMS coefficient $S$.

It should be noted that the presented formation here is only limited to dynamically formed BBHs as we have no primordial binaries in our simulations. Primordial binaries, especially massive binaries including BHs, are expected to impact the overall dynamical evolution of star clusters. However, Binaries reform dynamically on a relaxation timescale, and the majority of massive primordial binaries are disrupted within 100 Myr due to the dynamical interactions or kicks when compact objects are formed. As a result, the total number and properties of BH–BH binaries that remain within old star clusters are entirely determined by stellar dynamic processes \citep{Chatterjee2017}. If massive primordial binaries exist, the enhanced few-body interaction between massive binaries leads to the early escape of massive stars and BHs, which can impact the number of retained BH–BH binaries in a cluster \citep{Wang2022}.

\section{Summary and Conclusions}
\label{conclusion}

We studied the evolution of star clusters that could enter the DSC phase within a Hubble time. We conducted direct $N$-body simulations of 22 star cluster models, of which 10 clusters were not initially mass segregated ($S=0$) and the rest were entirely mass segregated ($S=1$). We assumed a canonical IMF with a natal kick of zero for SNe remnants. By analyzing the differences throughout their evolutionary process, we explored the formation and evolution of the DSC phase, the mass fraction of retained BHs in clusters at present, and the BH-BH binary formation rate.

We reasserted that the main differences in the energy generation from BHSub of star clusters with PMS coefficients of $S=0$ and $S=1$ can be attributed to three processes. First, in a cluster with a high PMS coefficient, low-mass stars have larger initial speeds compared to massive stars, as a result of being in the virial equilibrium. This enables them to travel further distances from the cluster core. In such regions, if low-mass stars are given a tiny boost in speed, the tidal force of the host galaxy can unbind them from the cluster. Second, S1 models have more massive stars at the centre. As a result, energy equipartition renders an outward direction for the energy flow from massive stars SNe (inner regions) to low-mass stars (outer regions). Third, three-body encounters are more likely when the BHSub is more compact, as is the case for S1 models.

We demonstrated that higher degrees of PMS make a cluster evolve dynamically faster, i.e. it expands to larger radii at a rapid pace, and its dissolution time shortens. Moreover, it also aids in retaining a large group of BHs inside the cluster. Clusters that lack natal kicks for their stellar remnants and reside in the innermost regions of a MW-like galaxy ($\RG\leq 4\kpc$), can reach the DSC phase within a Hubble time. For such clusters, the PMS affects the cluster's evolution in various fashions, as further discussed below

\begin{itemize}

\item S1 clusters has a dissolution time which is approximately $50$ per\,cent shorter than their S0 twin clusters. However, S1 clusters spend a larger fraction of their lifetime as a DSC, i.e. on average they have $\ttDSC$ of approximately twice as long as that of S0 clusters.

\item Highly mass-segregated clusters located within a Galactic orbital radius of $\leq48\kpc$ and initial densities (after gas expulsion) less than $\rho<\Msun[5]\pc^{-3}$ can ultimately become a DSC, whereas this is true for initially non-segregated clusters if they reside within $\RG\leq17\kpc$ and have a density of $\rho<1.6\times\Msun[4]\pc^{-3}$.

\item Clusters with different degrees of PMS, which are otherwise identical, have different expansion factors $\rh(\mathrm{max})/\rh(t=0)$, i.e. S1 models reach larger expansion factors. However, the difference becomes less in clusters with a higher retention fraction of BHs.

\item Our estimation suggests that approximately 24 per\,cent of the MW GCs may retain a population of BHs if they began with an exceptionally high $S$. However, this percentage decreases by half for clusters that were initially non-segregated (\citetalias{rostami2024}). Moreover, considering $S=1$, up to 14 per\,cent of the MW GCs exhibit a BH mass fraction exceeding 4 per\,cent and can turn into DSCs. This fraction diminishes to zero for the MW GCs with lower $S$ parameter.

\item Assuming that GCs form in small initial half-mass radius \citep{markskroupa} and adopting a canonical IMF, our models cannot explain the large fraction of BHs in clusters, as suggested by some studies, such as $\omega\,\mathrm{Cen}$. This is true, even if clusters were initially highly mass segregated. A larger initial $\rh$ or assuming a top-heavy IMF can help retain more black holes in clusters and increase the BH mass fraction to be consistent with previous studies (see \citetalias{rostami2024}).

\item The average rate of dynamical BBH formation, $\RBBH$, for S1 models is 2.5 times as large as that of S0 models. We noted that in S0 clusters $\RBBH$ asymptotically reaches zero, however, it levels off to a constant rate for S1 models. Interestingly, this increases the chance of observing X-rays and GWs in GCs.

\end{itemize}

Our findings aid us in gaining a better understanding of how GCs evolve in the presence of primordial mass segregation, as well as the observable properties of DSCs.

\section*{Data availability}
The data underlying this article are available in the article.

\bibliographystyle{mnras}
\bibliography{references} 






\bsp	
\label{lastpage}
\end{document}